\documentclass{article}

\pdfoutput=1

\PassOptionsToPackage{numbers, square}{natbib}

\usepackage[final]{neurips_2019}

\usepackage[utf8]{inputenc} 
\usepackage[T1]{fontenc}    
\usepackage{hyperref}       
\usepackage{url}            
\usepackage{booktabs}       
\usepackage{amsfonts}       
\usepackage{nicefrac}       
\usepackage{microtype}      

\usepackage{amsmath}
\usepackage{subcaption}
\usepackage{graphicx}
\usepackage{wrapfig}
\usepackage[ruled]{algorithm2e}

\title{Learned Interpolation for 3D Generation}
\date{}

\author{Austin Dill, Songwei Ge, Eunsu Kang, Chun-Liang Li\\
{\bf Barnabas Poczos} \\
Carnegie Mellon University\\
Pittsburgh, PA, United States \\
\{abdill, songweig, eunsuk, chunlial, bapoczos\}@andrew.cmu.edu\\
\newline
\newline
}

\begin{document}

\maketitle
\textit{}\vspace{-5em}
\section{Introduction}

Merging abstract concepts, also known as \textit{creative blending}, is frequently seen as a fundamental component of creativity, as this merging can allow novel concepts to emerge from simple components \cite{pereira2007creativity}. One computational way to approach this is \textit{interpolation}, a process that smoothly transitions from one instance to another.

In order to generate novel 3D shapes with machine learning, one must allow for such interpolations. The typical approach for incorporating this creative process is to interpolate in a learned latent space so as to avoid the problem of generating unrealistic instances by exploiting the model's learned structure. In 2D images, this often utilizes the trained Generative Adversarial Network~\cite{radford2015unsupervised,brock2018large} or Autoencoder~\cite{kingma2013auto,tolstikhin2017wasserstein}, which has shown promising results in creative generation~\cite{carter2017using}. As for the basic requirement, the process of the interpolation is supposed to form a semantically smooth morphing~\cite{berthelot2018understanding}. While this approach is sound for synthesizing realistic media such as lifelike portraits or new designs for everyday objects, it subjectively fails to directly model the unexpected, unrealistic, or creative \cite{DeepCloud, kingma2018glow}. 

In this work, we present a method for learning how to interpolate point clouds. By encoding prior knowledge about real-world objects, the intermediate forms are both realistic and unlike any existing forms. We show not only how this method can be used to generate "creative" point clouds, but how the method can also be leveraged to generate 3D models suitable for sculpture.  

\section{Interpolation as Generation}
\vspace{-1.5em}
\begin{figure}[h]
\centering
\includegraphics[width=\textwidth]{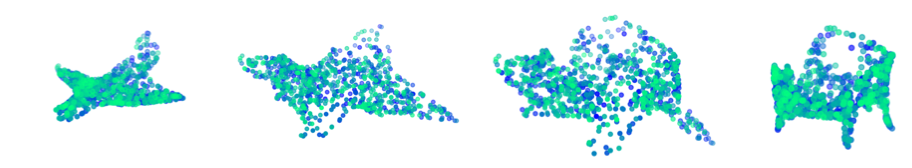}
\caption{Learned interpolation between an airplane and a chair.}
\label{learned_interp}
\end{figure}

\vspace{-1.5em}
\subsection{Naive Interpolation}

Consider two point clouds $X_a = \{p_i\}_{i=1}^n$ and $X_b = \{p'_i\}_{i=1}^n$, where each $p$ represent a 3-dimensional point. To generate a point cloud that semantically lies in between the inputs $X_a$ and $X_b$, a straightforward method is to generate a point cloud $X_{ab}$ as follows:

\begin{equation*}
    X_{ab} = \{ \alpha p_i + (1-\alpha) p'_i\}_{i=1}^n
\end{equation*}

\begin{wrapfigure}{R}{0.3\linewidth}
\vspace{-1.5em}
\centering
\includegraphics[width=0.3\textwidth]{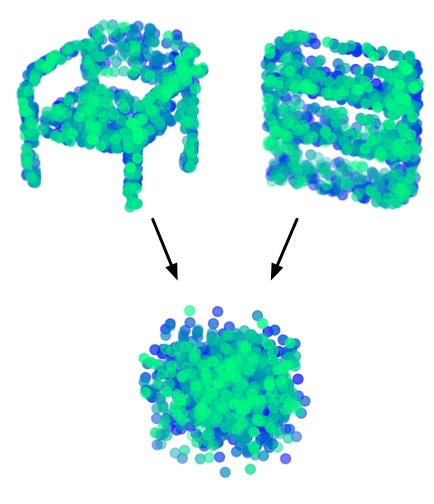}
\caption{Naive interpolation fails to produce an interesting midpoint.}
\label{naive_interp}
\end{wrapfigure}

Intuitively, this represents drawing a line between pairs of points in $X_a$ and $X_b$ and returning a points $\alpha$ percent of the distance between them. While this is simple to implement, it does not produce results that are semantically in between the objects the point clouds represent, as can be seen in Figure \ref{naive_interp}. 

\subsection{Learned Point Cloud Interpolation}

While prior algorithms for generating creative point clouds have relied on a pretrained model, our method directly learns a transformation from point clouds to point clouds by using interpolation as the guiding framework. We parameterize interpolation from a start point cloud $X_a$ to a goal point cloud $X_b$ where for each time step, we apply a learned transformation for each point independently, given an encoding of $X_b$, denoted by $E(X_b)$.

\begin{equation*}
    p^{t+1}_i = p^{t}_i + h_{t}(p^{t}_i, E(X_b))
\end{equation*}

In each of the above transformations, the function $h_t(\cdot)$ is a multilayer perceptron \cite{rosenblatt1958perceptron}. All of the transformations are trained to produce a set $\hat{X}_b$ after $T$ transformations so that $\hat{X}_b$ and $X_b$ are close in Chamfer Distance, an error metric frequently used with set generation tasks \cite{achlioptas2017learning}. The encoding network is parameterized as a Deep Sets model \cite{zaheer2017deep}. 

While this formulation allows us to visualize the trajectory of each point as it is transformed and allows us enough expressivity to approximate the goal point clouds, it does not enforce the requirement that each intermediate point cloud is realistic. For example it could allow a mapping to a completely meaningless intermediate state that would not be recognized as a plausible (if unusual) 3D object. 

For this reason we introduce an additional loss term motivated by computer graphics \cite{ezuz2018reversible}. 

\begin{equation*}
    \mathcal{L}(\hat{X}_b, X_b) = CD(\hat{X}_b, X^*) + \sum_i \sum_{j \in N(i)} (p_i - p_j)^2 - (\phi(p_i)-\phi(p_j))^2
\end{equation*}

With this added term, we are able to maintain the topology of the beginning object, causing the network to find the most plausible correspondence between the source object and the target function. This loss function only penalizes the output of the algorithm but has the side effect of ensuring each intermediate step is topologically consistent as well. 

\subsection{Mesh Generation}

\begin{figure}[!htb]
\minipage{0.24\textwidth}
\vspace{0.5em}
  \includegraphics[width=\linewidth]{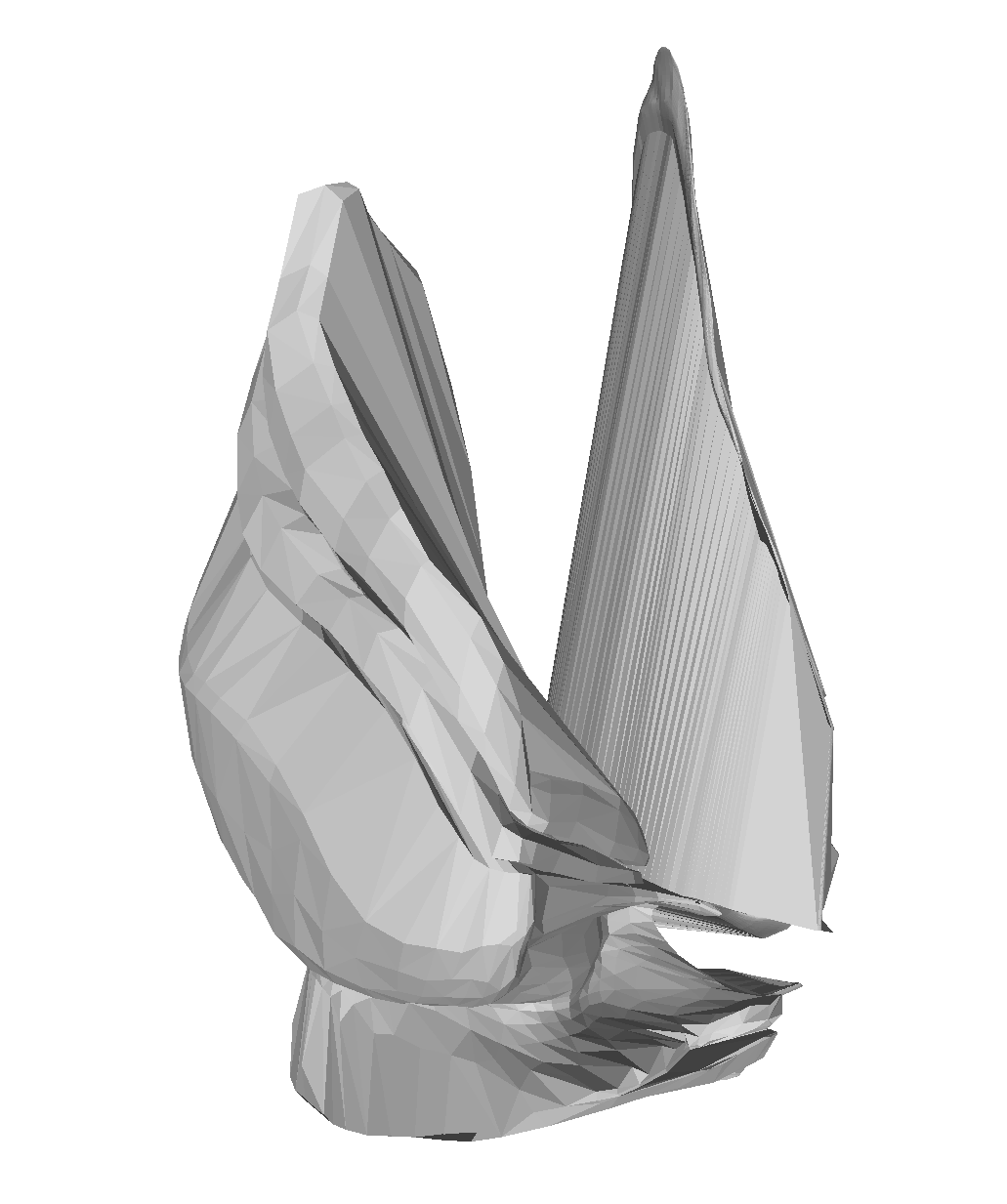}
\caption*{Toilet and Plant}
\endminipage\hfill
\minipage{0.24\textwidth}
\vspace{3.0em}
  \includegraphics[width=\linewidth]{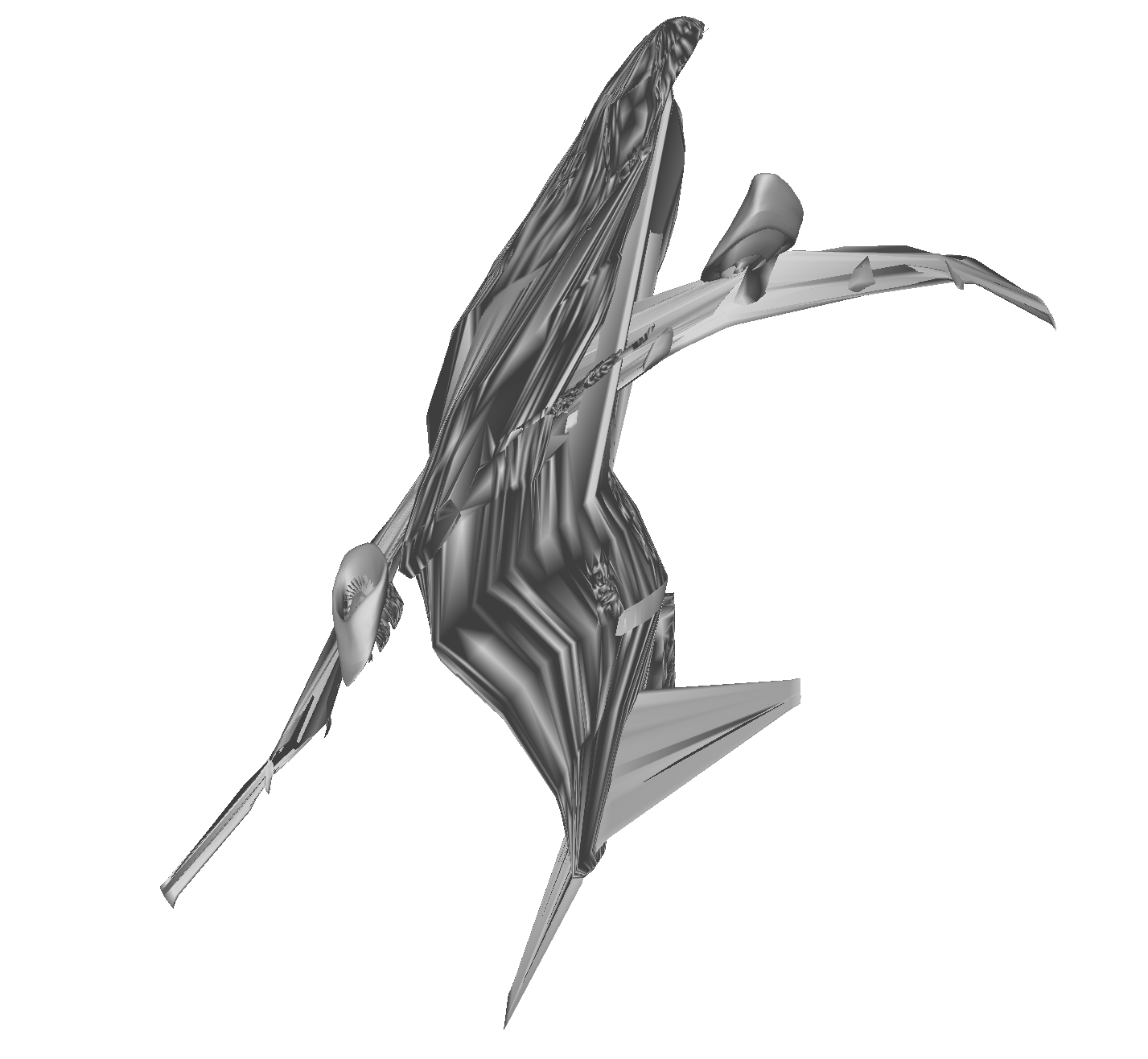}
 \caption*{Airplane and Person}
\endminipage\hfill
\minipage{0.24\textwidth}%
\vspace{4em}
  \includegraphics[width=\linewidth]{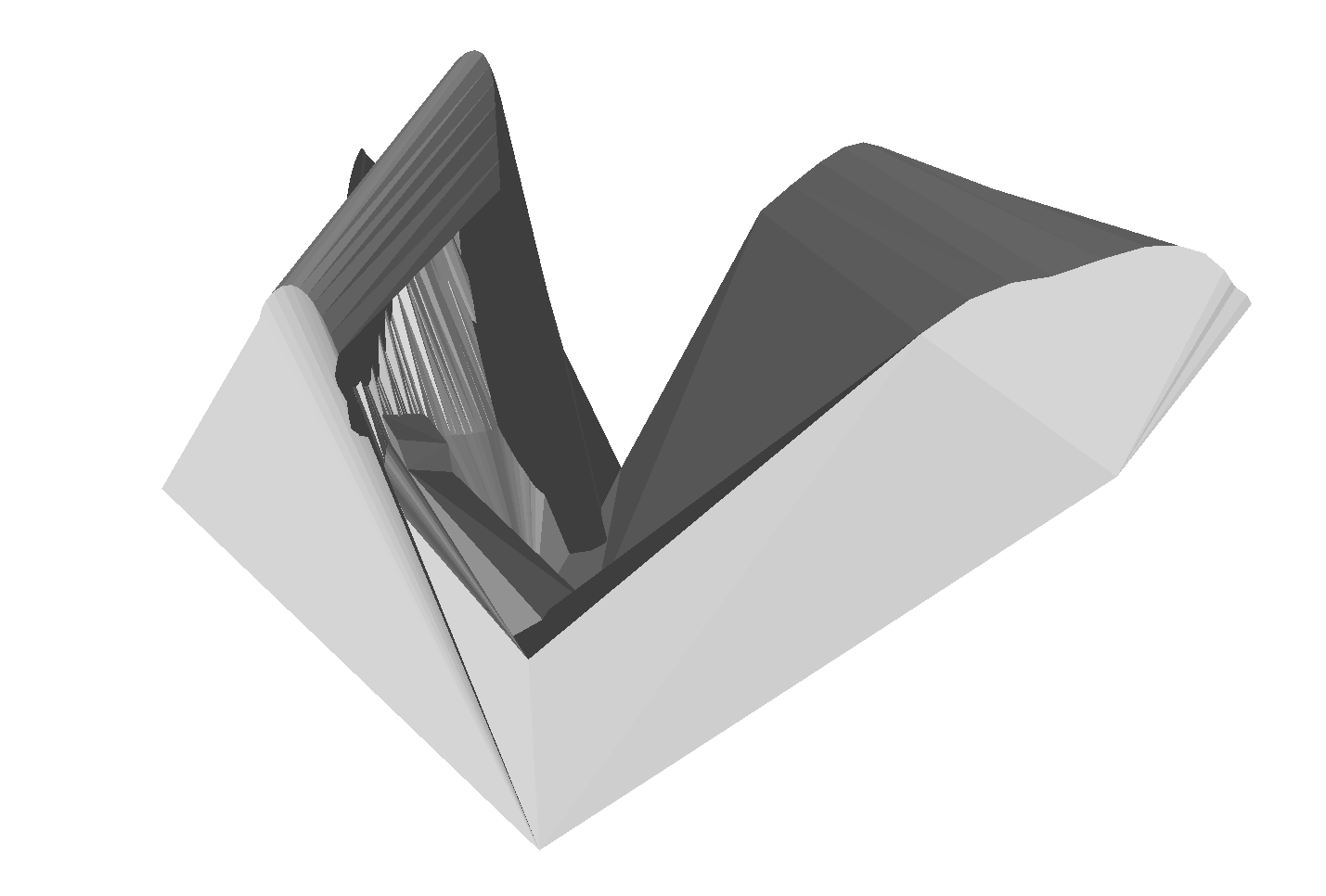}
 \vspace{0.5em}
\caption*{Piano and Bowl}
\endminipage\hfill
\minipage{0.24\textwidth}%
  \includegraphics[width=\linewidth]{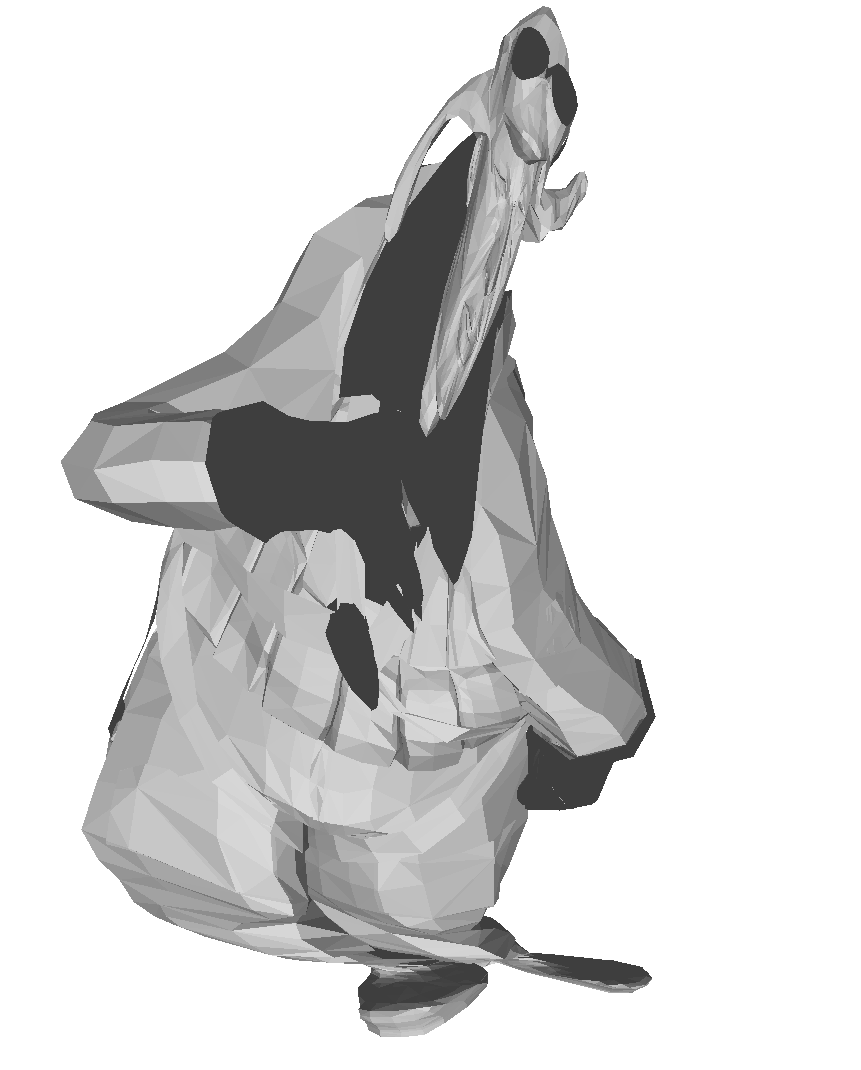}
\caption*{Person and Plant}
\endminipage
\caption{Generated meshes from interpolating between pairs of objects.}
\end{figure}

This technique allows one to use the vertices from a mesh as input, providing us with the correspondence needed to mesh the output. This removes the problem of meshing for creative sculpture generation, the motivating factor for creative sculpture generating algorithms \cite{ge2019developing}. In the author's opinion, the conflict between the representational advantage of point clouds for machine learning tools and the artist's frequent need for solid 3D shapes has limited the adoption of generative models for sculptural art. Our method therefore represents an important step forward for creative AI. 
\clearpage

\small

\bibliographystyle{abbrvnat}
\bibliography{references}

\end{document}